\documentclass[twocolumn]{aastex631}

\usepackage{natbib}
\usepackage{hyperref}

\usepackage{color}
\definecolor{black}{rgb}{0,0,0}
\definecolor{red}{rgb}{1,0,0}

\shorttitle{AASTeX v6.3.1 Sample article}
\shortauthors{Sicardy et al.}
\graphicspath{{./}{figures/}}
\begin{document}

\title{%
Pluto's atmosphere in plateau phase since 2015 from a stellar occultation at Devasthal
}%


\author[0000-0003-1995-0842]{Bruno Sicardy}
\affiliation{%
LESIA, Observatoire de Paris, \\
Universit\'e PSL, CNRS, Sorbonne Universit\'e, \\
Universit\'e de Paris, 5 place Jules Janssen, 92195 Meudon, France
}

\author{Nagarhalli M. Ashok}
\affiliation{Physical Research Laboratory, Ahmedabad, 380009, Gujarat, India \\
}

\author[0000-0001-5917-5751]{Anandmayee Tej}
\affiliation{Indian Institute of Space Science and Technology, Thiruvananthapuram, 695547, Kerala, India}

\author{Ganesh Pawar}
\author{Shishir Deshmukh}
\author{Ameya Deshpande}
\affiliation{Akashmitra Mandal, Kalyan, 421301, Maharashtra, India}

\author{Saurabh Sharma}
\affiliation{Aryabhatta Research Institute of Observational Sciences, Manora Peak, Nainital 263002, India}

\author[0000-0002-2193-8204]{Josselin Desmars}
\affiliation{Institut Polytechnique des Sciences Avanc\'ees IPSA, 94200 Ivry-sur-Seine, France}
\affiliation{IMCCE, Observatoire de Paris, \\
PSL Research University, CNRS, Sorbonne Universit\'e, Univ. Lille, 75014 Paris, France
}%

\author[0000-0002-8211-0777]{Marcelo Assafin}
\affiliation{%
Universidade Federal do Rio de Janeiro - Observat\'orio do Valongo, Ladeira Pedro Ant\^onio 43, CEP 20.080-090 Rio de Janeiro - RJ, Brazil
}%
\affiliation{%
Laborat\'orio Interinstitucional de e-Astronomia - LIneA, Rua Gal. 4 Jos\'e Cristino 77, Rio de Janeiro 20921-400, Brazil
}%

\author[0000-0002-8690-2413]{Jose Luis Ortiz}
\affiliation{%
Instituto de Astrof\'{\i}sica de Andaluc\'{\i}a, IAA-CSIC, Glorieta de la Astronom\'{\i}a s/n, 18008 Granada, Spain
}%

\author[0000-0002-4106-476X]{Gustavo Benedetti-Rossi}
\affiliation{%
LESIA, Observatoire de Paris, \\
Universit\'e PSL, CNRS, Sorbonne Universit\'e, \\
Universit\'e de Paris, 5 place Jules Janssen, 92195 Meudon, France
}
\affiliation{%
UNESP - S\~ao Paulo State University, Grupo de Din\^amica Orbital e Planetologia, CEP 12516-410, Guaratinguet\'a, SP, Brazil
}%
\affiliation{%
Laborat\'orio Interinstitucional de e-Astronomia - LIneA, Rua Gal. 4 Jos\'e Cristino 77, Rio de Janeiro 20921-400, Brazil
}%

\author[0000-0003-2311-2438]{Felipe Braga-Ribas}
\affiliation{%
Federal University of Technology-Paran\'a (UTFPR / DAFIS), Curitiba, Brazil
}%
\affiliation{%
Observat\'orio Nacional/MCTIC, Rio de Janeiro, Brazil
}%
\affiliation{%
Laborat\'orio Interinstitucional de e-Astronomia - LIneA, Rua Gal. 4 Jos\'e Cristino 77, Rio de Janeiro 20921-400, Brazil
}%

\author[0000-0003-1690-5704]{Roberto Vieira-Martins}
\affiliation{%
Observat\'orio Nacional/MCTIC, Rio de Janeiro, Brazil
}%
\affiliation{%
Laborat\'orio Interinstitucional de e-Astronomia - LIneA, Rua Gal. 4 Jos\'e Cristino 77, Rio de Janeiro 20921-400, Brazil
}%
\affiliation{%
Universidade Federal do Rio de Janeiro - Observat\'orio do Valongo, Ladeira Pedro Ant\^onio 43, CEP 20.080-090 Rio de Janeiro - RJ, Brazil
}%

\author[0000-0002-1123-983X]{Pablo Santos-Sanz}
\affiliation{%
Instituto de Astrof\'{\i}sica de Andaluc\'{\i}a, IAA-CSIC, Glorieta de la Astronom\'{\i}a s/n, 18008 Granada, Spain
}%

\author{Krishan Chand}
\affiliation{Aryabhatta Research Institute of Observational Sciences, Manora Peak, Nainital 263002, India}

\author{Bhuwan C. Bhatt}
\affiliation{Indian Institute of Astrophysics, II Block, Koramangala, Bangalore, 560034, India}

\begin{abstract}
A stellar occultation by Pluto was observed on 6 June 2020
with the 1.3-m and 3.6-m telescopes located at Devasthal, Nainital, India,
using imaging systems in the I and H bands, respectively.
From this event, we derive a surface pressure for Pluto's atmosphere of
$p_{\rm surf}= 12.23^{+0.65}_{-0.38} $~$\mu$bar.
This shows that Pluto's atmosphere is in a plateau phase since mid-2015, a result 
which is in excellent agreement with the Pluto volatile transport model of \cite{mez19}.
This value does not support the pressure decrease reported by independent teams, 
based on occultations observed in 2018 and 2019,
see \cite{you21} and \cite{ari20}, respectively.
%
\end{abstract}
\section{Introduction} \label{sec:intro}

Owing to its high obliquity (120$^\circ$) and high orbital eccentricity (0.25), 
Pluto suffers intense seasonal episodes. 
Its poles remain, for decades, in permanent sunlight or darkness over its 248-year heliocentric revolution.
This leads to strong effects on its N$_2$ atmosphere that is mainly controlled 
by vapor pressure equilibrium with the surface N$_2$ ice.
The NASA New Horizons flyby in July 2015 revealed a large depression, Sputnik Planitia, filled by N$_2$ ice
\citep{ste15}, which appeared to be the main engine that controls the seasonal
variation of atmospheric pressure during one seasonal cycle \citep{ber16,ber18,joh21b}.
%
Apart from these crucial results, a comprehensive review 
of the composition, photochemistry, atmospheric dynamics, circulation and escape processes 
derived from the New Horizons data is presented in \cite{gla19}. 

In parallel, ground-based observations of stellar occultations  
allowed various teams to accurately monitor Pluto's atmospheric pressure since 1988. 
A compilation of twelve occultations observed between 1988 and 2016 shows 
a three-fold monotonic increase of pressure during this period, that can be explained
by the progression of summer over the northern hemisphere, exposing Sputnik Planitia to the solar radiation \citep{mez19}.
This increase can be explained consistently by a Pluto volatile transport model, 
which predicts that the pressure should peak around 2020. 
A gradual decline should then last for two centuries under the combined effects of Pluto's recession 
from the Sun and the prevalence of the winter season over Sputnik Planitia. 

Here we present the results of a stellar occultation by Pluto that occurred on 6 June 2020. 
It was observed in the near-infrared
by two large telescopes at the Devasthal station, Nainital, India. 
The high signal-to-noise ratio light curves obtained with these instruments allow us to derive an
accurate value of Pluto's atmospheric pressure, using the same approach as in 
\cite{dia15} and \cite{mez19}.
This occultation was particularly timely as it can test the validity of 
the current models of Pluto's atmosphere evolution.
Moreover, as Pluto is now moving away from the Galactic plane as seen from Earth, 
stellar occultations by the dwarf planet are becoming increasingly rare, making this event a decisive one.

\section{Observations and Data Analysis}

\subsection{Occultation}

The 6 June 2020 occultation campaign was organized within the Lucky Star 
project\footnote{https://lesia.obspm.fr/lucky-star/}.
The prediction used the Gaia DR2 position at epoch of occultation (Table~\ref{tab_result})
and the NIMAv8/PLU055 Pluto's ephemeris derived from previous occultations observed since 
1988 \citep{des19}.
More information on the event (shadow path, charts, photometry, etc.) is available in a dedicated web page\footnote{https://lesia.obspm.fr/lucky-star/occ.php?p=31928}. 

The event was successfully recorded in the I and H bands using
the 1.3-m Devasthal Fast Optical Telescope (DFOT) and 
the 3.6-m Devasthal Optical Telescope (DOT), respectively. 
The I and H band magnitudes of the occulted star are $\sim$12.3 and 11.6, respectively, while that of Pluto during the epoch of occultation were $\sim$13.8 and 13.3, respectively. 
%
Thus, the occulted star was more than 1.5 mag brighter than the combined Pluto + Charon system, ensuring a high contrast event,
i.e. a significant drop of the total flux, 
which combines the fluxes from the star, Pluto and Charon.

Both telescopes 
are operated by the Aryabhatta Research Institute of Observational Sciences (ARIES) located at Nainital, India. Observations were also planned with the 2-m Himalayan Chandra Telescope (HCT), Hanle, 
operated by the Indian Institute of Astrophysics, Bangalore, India. However, the event was clouded out at that site.
%
%
%

The event was observed in the I-band  with DFOT-Andor DZ436 camera (2048$\times$2048-pixel; plate scale $\sim$0.5~arcsec/pixel).
The central 401$\times$401 pixel region was used in 2$\times$2 binning mode. 
With a readout rate of 1~MHz, shift speed of 16~$\mu$s and exposure time of 1.7-s, 
a total cycle time of 2.507-s was achieved. 
The final acquired image was a {\tt FITS} data cube of 600 frames.
A simultaneous observation was carried out with DOT in the H-band with the TIRCAM2 instrument \citep{{nai12},{bau18}},
an imaging camera housing 
a cryo-cooled Raytheon InSb Aladdin III Quadrant focal plane infrared array 
(512$\times$512-pixel; plate scale = 0.167~arcsec/pixel).
The full frame was used with a readout rate of 1~MHz, exposure time of 5~s, and total cycle time of 5.336~s. 
The final acquired image was a FITS data cube of 255 frames.

\begin{figure*}
\centering
\includegraphics[scale=0.5]{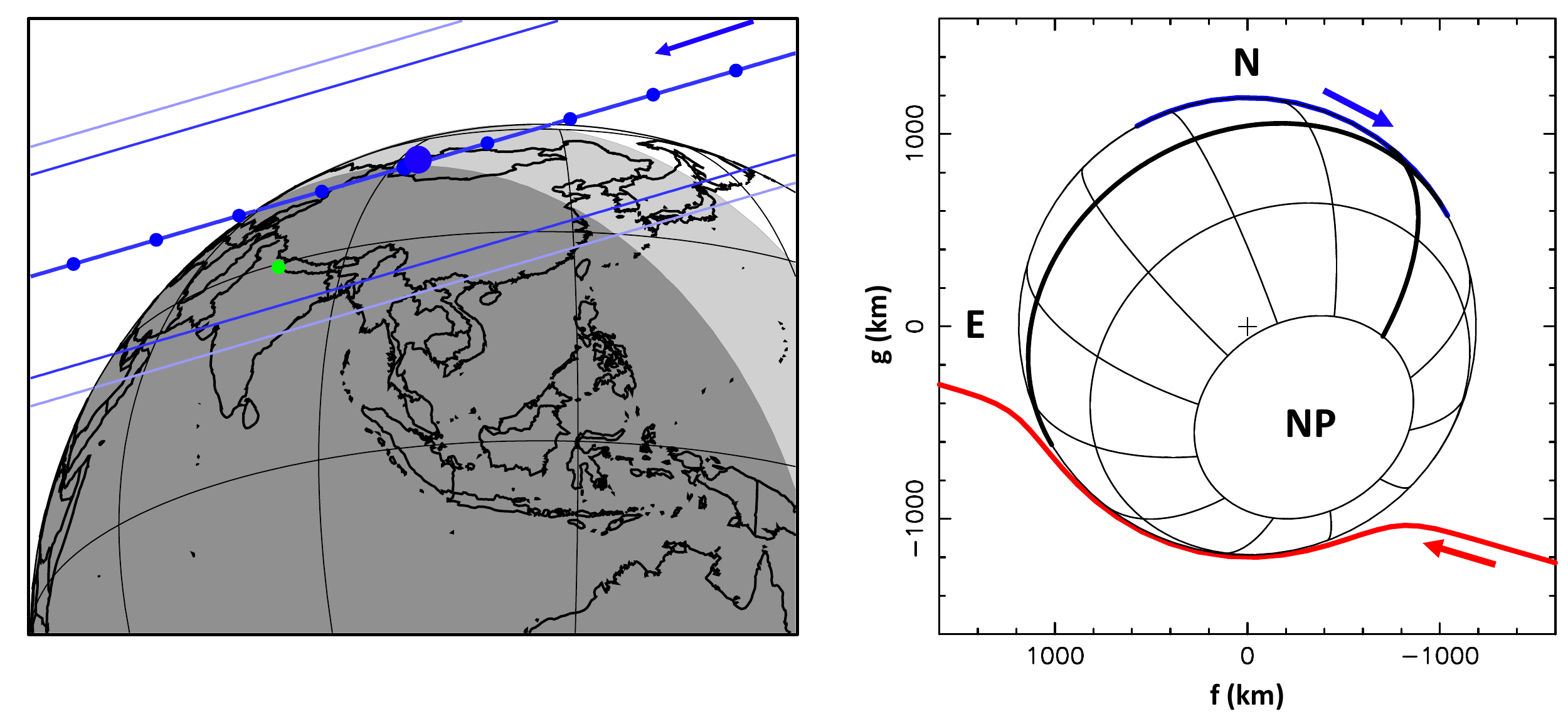}
\caption{%
{\it Left:} 
The reconstructed shadow path of Pluto on 6 June 2020.
The blue dots on the shadow central line are plotted every minute.
The larger one marks the Geocentric closest approach at 19:01:01.7 UT, 
and the arrow indicates the direction of motion.
The green dot is Devasthal's position.
The dark and light blue lines on each side of centrality correspond to the stellar half-light level and 
1\% stellar drop level (the practical detection limit), respectively.
The dark gray region is for astronomical night (Sun more than 18 deg below the horizon), while the
light gray region is for astronomical twilight (Sun between 0 and 18 deg below the horizon).
{\it Right:}
Geometry of the 6 June 2020 stellar occultation event.
Pluto's orientation is shown at 19:02 UT. 
The thicker lines are the equator and the prime (Charon-facing) meridians, respectively.
The letters N, E and NP indicate the celestial north, celestial east and Pluto's north pole, respectively.
The red and the blue lines represent, respectively, the motion of 
the primary and secondary stellar images relative to Pluto, as seen from Devasthal, see text for details.
%
The quantities $f$ and $g$ are the offsets of the stellar images with respect to Pluto's center, 
marked with a `+' symbol. 
}%
\label{fig_shadow_path}
\end{figure*}

From the light curve fitting described below, we reconstructed Pluto's shadow path on Earth and 
the geometry of the occultation 
(Fig.~\ref{fig_shadow_path}).
Note in this figure that two stellar images (primary and secondary, see \citealt{sic16}) actually scanned Pluto's limb. 
However, the flux of the secondary image was always 
fainter than that of the primary by a factor larger than 25, making it negligible in our 
case.
Consequently, this event essentially scanned 
the northern, summer hemisphere of Pluto. 

\begin{deluxetable*}{ll}
\tablecaption{%
Circumstances of observations, adopted parameters and result of the atmospheric fit.
\label{tab_result}
}%
\tablewidth{0pt}
\tablehead{\multicolumn{2}{c}{Observation log}}
\startdata
\hline
\multicolumn{2}{c}{3.6-m (DOT)} \\
\hline
Coordinates, altitude & 79$^{\circ}$ 41$^{\prime}$ 3.6$^{\prime\prime}$ E, 29$^{\circ}$ 21$^{\prime}$ 39.4$^{\prime\prime}$ N, 2450 m \\
Camera 								& TIRCAM2 Raytheon InSb array \\
Filter ($\lambda/\Delta \lambda$, $\mu$m)	& H (1.60/0.30) \\
Exposure time/Cycle time (s)               		& 5./5.336 \\                               
\hline
\multicolumn{2}{c}{1.3-m (DFOT)} \\
\hline
Coordinates, altitude   & 79$^{\circ}$ 41$^{\prime}$ 6.1$^{\prime\prime}$ E, 29$^{\circ}$ 21$^{\prime}$ 41.5$^{\prime\prime}$ N,  2450 m \\
Camera 		            & ANDOR DZ436 \\
Filter ($\lambda/\Delta \lambda$, $\mu$m)	&   I (0.85/0.15) \\
Exposure time/Cycle time (s)               	&  1.7/2.507 \\                               
\hline
\multicolumn{2}{c}{Occulted star} \\
\hline
Identification (Gaia DR2)       & 6864932072159710592 \\
J2000 position at epoch (ICRF)  &  $\alpha= 19^{\rm h} 45^{\rm m} 33.9079^{\rm s}$, $\delta= -22^{\circ} 10^{\prime} 19.128^{\prime\prime}$ \\
\hline
\multicolumn{2}{c}{Pluto's body} \\
\hline
Mass$^1$            & $GM_P=  8.696 \times 10^{11}$ m$^3$ sec$^{-2}$ \\
Radius$^1$          & $R_P= 1187$~km \\
Geocentric distance & $4.97407 \times 10^9$ km \\
\hline
\multicolumn{2}{c}{Pluto's atmosphere} \\
\hline
N$_2$ molecular mass    & $\mu= 4.652 \times 10^{-26}$ kg \\
N$_2$ molecular         & $K = 1.091 \times 10^{-23}$ \\ 
Refractivity$^2$        &  $+6.282 \times 10^{-26}/\lambda_{\rm \mu m}^2)$ cm$^3$ molecule$^{-1}$   \\
Boltzmann constant      & $k= 1.380626 \times 10^{-23}$ J K$^{-1}$ \\
\hline
\multicolumn{2}{c}{Results of atmospheric fit (with 1$\sigma$ error bars) } \\
\hline
Pressure at radius 1215 km                                      & $p_{\rm 1215}= 6.655^{+0.35}_{-0.21} $~$\mu$bar \\
Surface pressure$^3$                                            & $p_{\rm surf}= 12.23^{+0.65}_{-0.38} $~$\mu$bar \\
Closest approach distance of Devasthal to shadow center$^4$   	& $\rho_{\rm C/A, D}= -735^{+7}_{-15}$ km \\
Closest approach time of Devasthal to shadow center             & $t_{\rm C/A, D}$= 19:02:43.0$\pm$0.14 s UT \\
Geocentric closest approach distance to shadow center$^4$   	& $\rho_{\rm C/A, G}= +6044^{+15}_{-7}$ km \\
Geocentric closest approach time to shadow center$^5$       	& $t_{\rm C/A, G}$= 19:01:01.7$\pm$0.14 s UT \\
\enddata
\tablecomments{%
$^1$\cite{ste15}, where $G$ is the constant of gravitation.
$^2$\cite{was30}, where $\lambda_{\rm \mu m}$ is the wavelength expressed in microns.
$^3$Using a ratio $p_{\rm surf}/p_{\rm 1215}=1.837$ given by the template model \cite{mez19}.
$^4$Negative (resp. positive) values mean that the point considered went south (resp. north) of the shadow center.
$^5$Although the quoted error bar is small, a systematic error of about 1~s may be present
in $t_{\rm C/A, G}$, see text.
}%
\end{deluxetable*}

\subsection{Light curve fitting}
The observed datasets were fitted with synthetic light curves
using the method described in \cite{dia15},
in particular with the same template temperature profile, $T(r)$, where $r$ is the distance to Pluto's center. 
The approach involves the simultaneous fitting of the observed refractive occultation light curves by 
synthetic profiles that are generated by a ray-tracing code that uses the Snell–Descartes law.
The various parameters used in our fitting procedure are listed in Table~\ref{tab_result}.

\begin{figure*}
\centering
\includegraphics[scale=0.4]{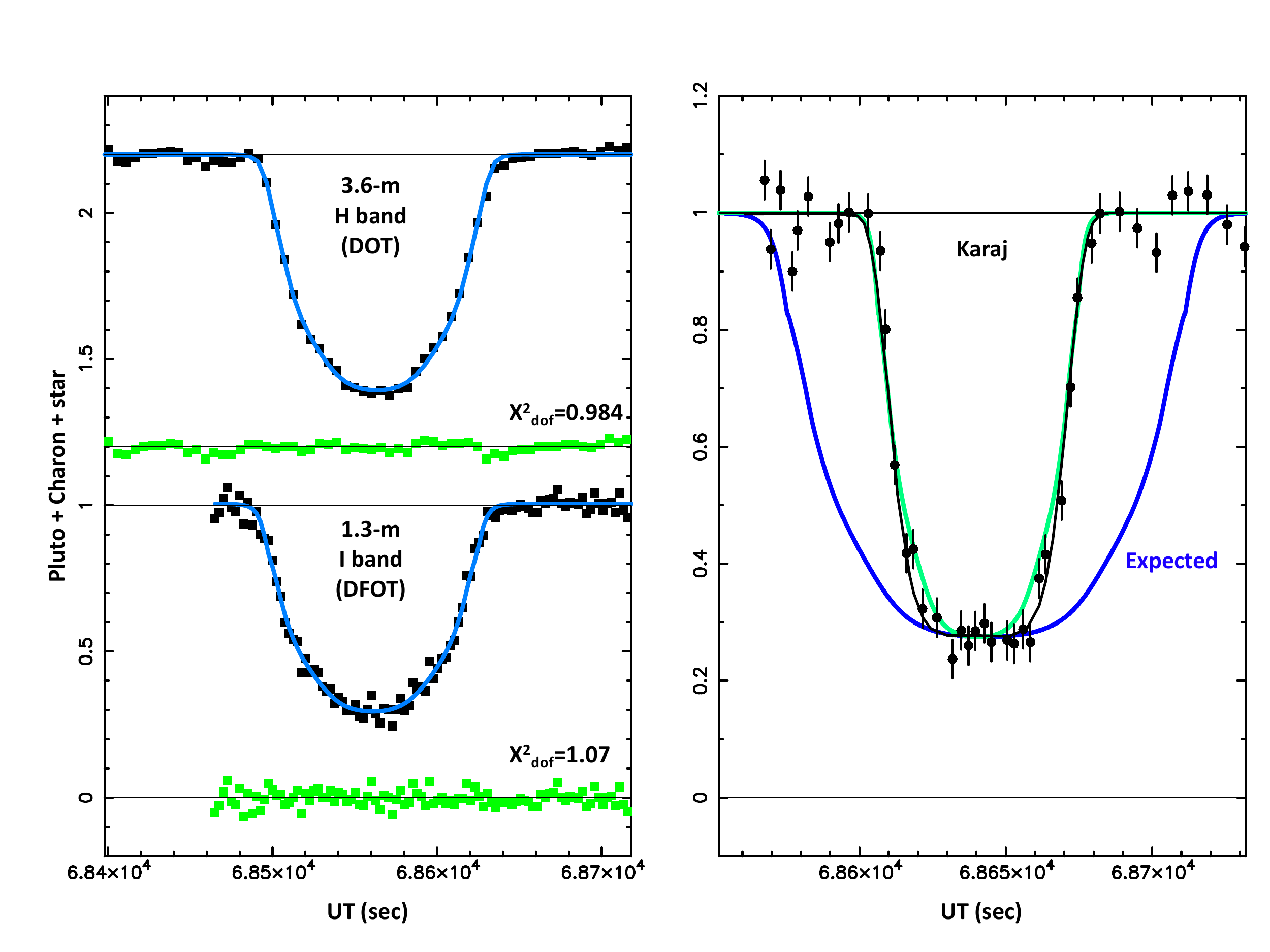}
\caption{%
{\it Left:} 
The blue curves are a simultaneous fit to our 6 June 2020 Pluto occultation 
light curves (black squares) obtained with the 3.6-m and 1.3-m telescopes of ARIES at Desvasthal, 
over a 320-s interval bracketing the event.
The residuals (observation-minus-model) are plotted in green below each light curve.
The parameters of the best atmospheric model are listed in Table~\ref{tab_result}.
The value of $\chi^2_{\rm dof}$, the $\chi^2$  per degree of freedom for each  fit,
is displayed at the lower right corner of each light curve. 
The lower and upper horizontal lines are the normalized total flux (star+Pluto+Charon) 
and the zero flux levels, respectively.
The 3.6-m light curve has been shifted vertically by +1.2 for better viewing.
{\it Right:}
Black bullets are
the data points obtained at the Karaj station during the same occultation event, 
and the black line is the associated best fitting model 
from \cite{por21} (credit: A\&A 653, L7, 2021, reproduced by permission \copyright ESO).
The blue curve is our expected light curve at Karaj using the results of \cite{por21}, 
i.e. $p_{\rm surf}= 12.36$~$\mu$bar and a
closest approach distance to Pluto's shadow at that station of 605.3~km.
It shows a large discrepancy by a factor of about two in the time scale when compared 
with the expected light curve.
Assuming a timing problem at Karaj, and trying to superimpose our synthetic curves, 
we obtain a significant discrepancy at the bottom of the occultation light curve 
between our model (green curve) and and the model of \cite{por21}.
This reveals an inconsistency between the ray tracing approach adopted by us and by \cite{por21}.
%
}%
\label{fig_fit}
\end{figure*}

There are $M=4$ adjusted parameters in our model: 
(1) $p_{\rm surf}$, the pressure at Pluto's surface, 
(2) $\Delta \rho$, the cross-track offset to Pluto's ephemeris, 
and 
(3-4) the two Pluto+Charon's contributions $\phi_0$ to each of the two ARIES light curves. 
Owing to less than optimal sky conditions prevailing, observations to separately measure the occulted star and Pluto's system the nights prior to or after the event was not possible. Due to this unavailability of calibration data, the $\phi_0$'s in our analysis are not known.

There is a fifth adjusted parameter which is completely uncorrelated with the other four, the 
time shift, $\Delta t$, that needs to be applied to Pluto's ephemeris to best fit the data. 
This parameter accounts for the ephemeris offset to apply along Pluto's apparent motion and for errors 
in the star position.
It finally provides, $t_{\rm C/A, G}$,  the time of closest approach of Pluto to the star 
in the sky plane, as seen from the Geocenter, see Table~\ref{tab_result}.

When fitting the two Devasthal light curves, a discrepancy of 2.4~s appeared in the best-fitting
$\Delta t$ derived from each telescope, thus revealing a problem in the recording of the absolute times at one (or both) telescopes. 
%
%
Since it is difficult to decipher the origin of this discrepancy, any attempt to correct for the same would be futile.
Hence, 
we have chosen to apply independent $\Delta t$
to each telescope, and calculate the final $t_{\rm C/A, G}$ as an average of the two obtained values,
weighted by the quality of the two light curves (measured by the noise in the data).
With this approach, although the internal error bar on $t_{\rm C/A, G}$ is small 
($\pm$0.14~s, Table~\ref{tab_result}), a systematic uncertainty of the order of 1~s still remains in
the quoted value of $t_{\rm C/A, G}$.

The best fit to the data are shown in Fig~\ref{fig_fit}. 
The function $\chi^2~=~\sum_1^N~[(\phi_{\rm i,obs}-\phi_{\rm i,syn})/\sigma_i]^2$ 
was used to assess the quality of the fit, 
where $\sigma_i$ reflects the noise level of each of the $N=362$ data points, and
$\phi_{\rm i,obs}$ and $\phi_{\rm i,syn}$ are the observed and 
synthetic fluxes at the $i^{\rm th}$ data point, respectively.
Satisfactory fits are obtained for a $\chi^2$ value per degree of freedom 
$\chi^2_{\rm dof}= \chi^2_{\rm min}/(N-M) \sim 1$,
where $\chi^2_{\rm min}$ is the minimum value of $\chi^2$ obtained in the fitting procedure.
This is the case here, with individual values 
$\chi^2_{\rm dof}=0.984$ and $\chi^2_{\rm dof}=1.07$ at DOT and DFOT, respectively, 
and a global value 
$\chi^2_{\rm dof}=1.03$ corresponding  to the simultaneous fit to the two light curves.

Our fit is mainly sensitive to regions above 30~km altitude, 
so our primary result is the pressure near the radius 1215~km, 
$p_{1215}= 6.665^{+0.35}_{-0.21}$~$\mu$bar (Table~\ref{tab_result}).
A factor of 1.837 is then applied to convert this into $p_{\rm surf}$, see \cite{mez19}.
Fig.~\ref{fig_t_p} shows the 
$\chi^2$ map plotted as a function of the two parameters $p_{\rm surf}$ and $\Delta \rho$.
Note that because there is only one occultation chord at hand (Fig.~\ref{fig_shadow_path}), 
a correlation between $p_{\rm surf}$ and $\Delta \rho$ is observed. 
Using the $\chi_{\rm min}^2+1$ criterion, we obtain the best-fitting value of 
$p_{\rm surf}= 12.23^{+0.65}_{-0.38} $~$\mu$bar. 
The marginal error bar quoted here is estimated by ignoring the value of $\Delta \rho$.

Besides the observations presented here, the 6 June 2020 event was observed by an independent team, see \cite{por21}.
The event was observed at low elevation (6 degrees) near the city of Karaj in Iran with a 60-cm telescope. 
The authors mention that they use the ray-tracing method of \cite{dia15}, i.e.
a procedure that should be fully consistent with our own approach. 

However, their derived value $p_{\rm surf}= 12.36 \pm 0.38$~$\mu$bar, is questionable.
First, the time axis for the Karaj light curve in \cite{por21} is 
wrong by a large factor of about two,
see Fig.~\ref{fig_fit}. This makes impossible to obtain any realistic value of $p_{1215}$ from this light curve.
Secondly, assuming that the Karaj time axis has some (undocumented) problems, 
considering the occultation geometry at that station 
and adopting the $p_{1215}= 6.72$~$\mu$bar value of \cite{por21},
we generated a synthetic model using our own ray-tracing code.
By shrinking the time scale in an attempt to superimpose 
our model (green curve in Fig.~\ref{fig_fit}) onto the results of \cite{por21}
(black curve), we see a clear discrepancy between our models 
in the deepest part of the occultation.
This shows that the ray tracing code of \cite{por21} is inconsistent with ours.
Consequently, the results of \cite{por21} are impossible to obtain considering their published light curve,
and probably stems from improper use of their ray-tracing code.
%
Finally, we note that the error bar for $p_{\rm surf}$ is inconsistent with the error bar that the
authors obtain for the pressure at radius 1215~km, 
$p_{1215}= 6.72 \pm 0.48$~$\mu$bar, see their Fig.~3. 
As the error bar scales like the value of the pressure,
the error bar for the surface pressure should be $\pm 0.88$~$\mu$bar, not $\pm 0.38$~$\mu$bar.


\section{Pressure evolution}

\begin{figure*}
\centering
\includegraphics[scale=0.25]{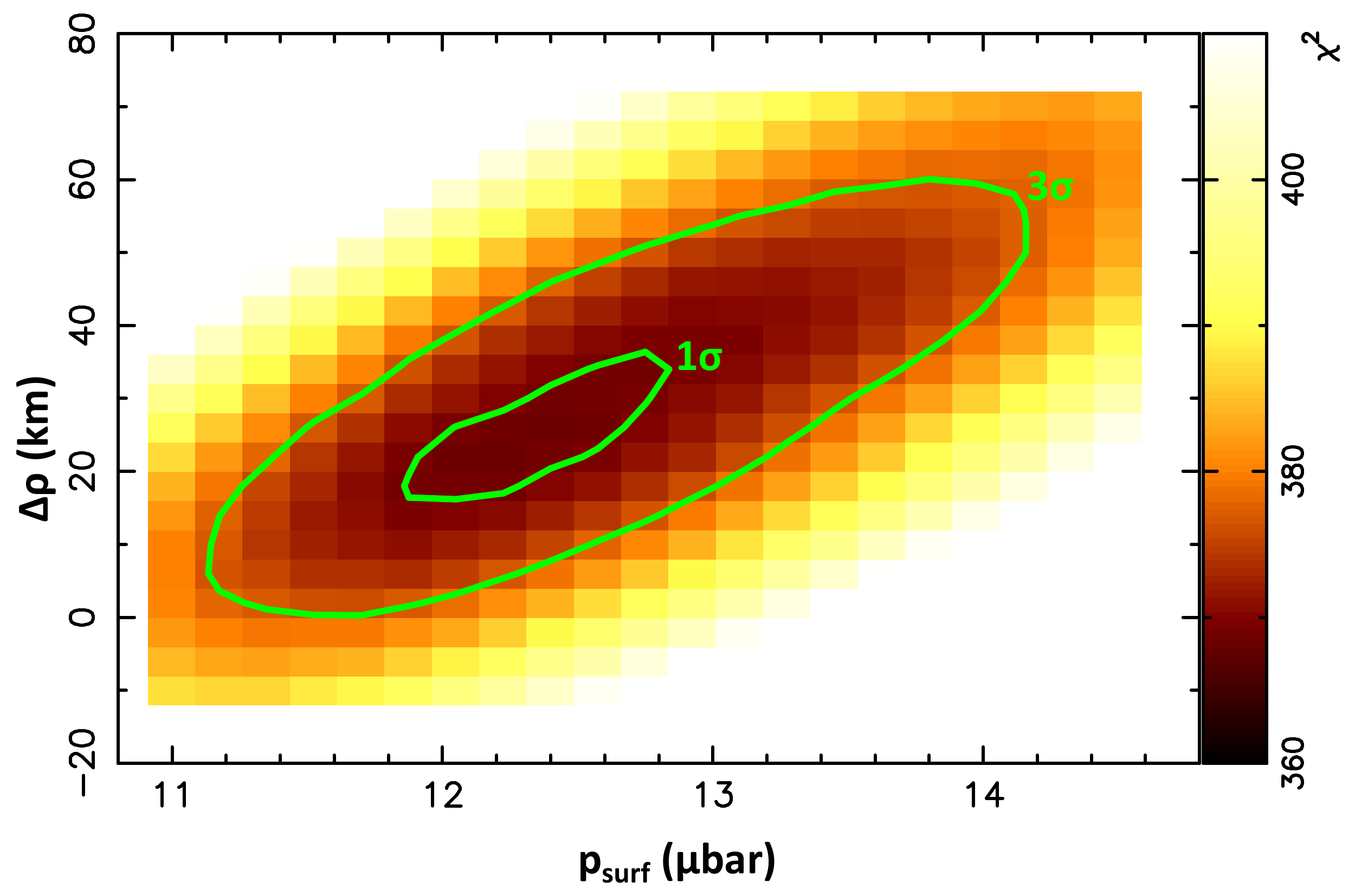}
\includegraphics[scale=0.40]{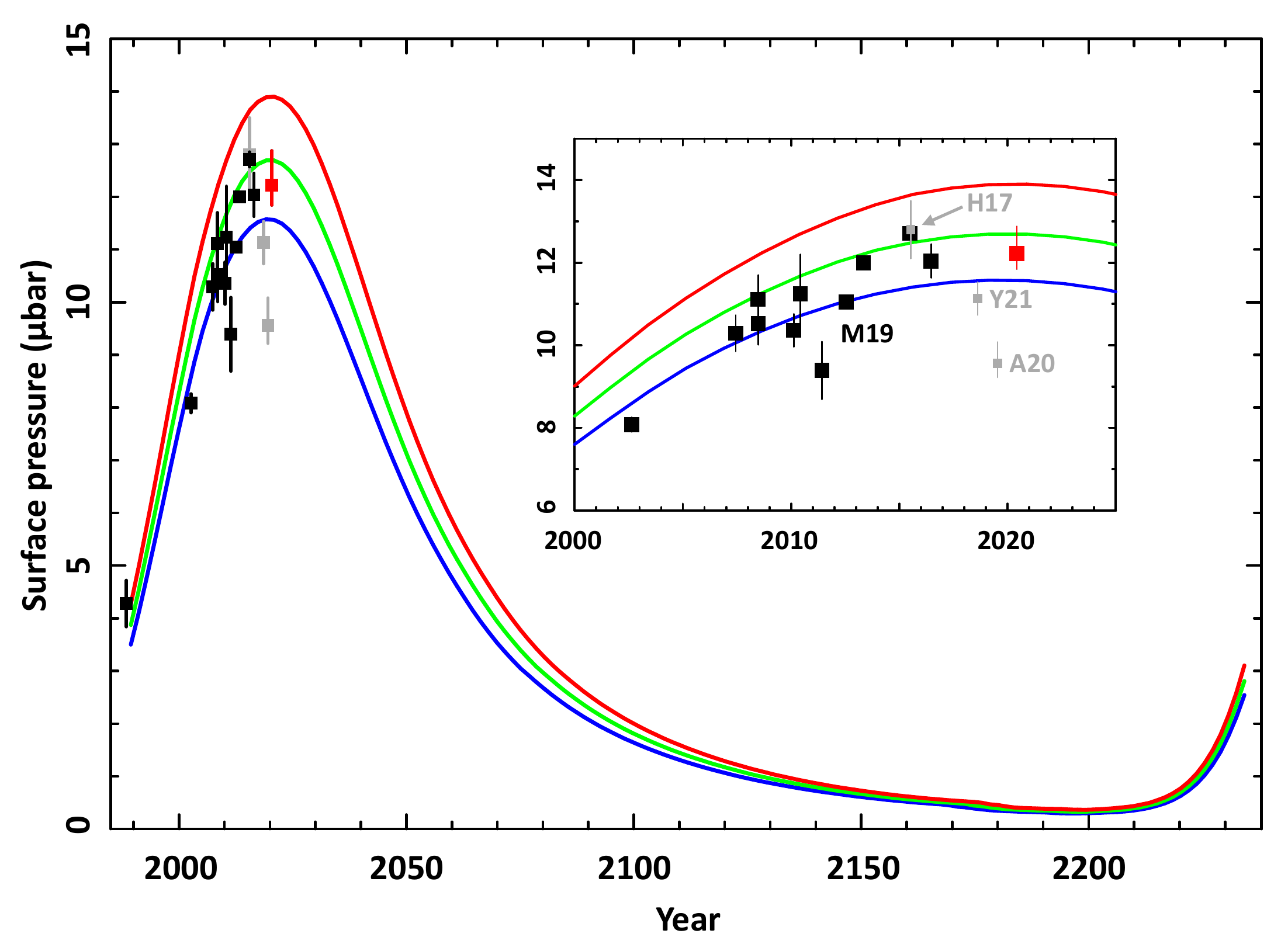}
\caption{%
{\it Left:} 
The $\chi^2$ map used to derived the best-fitting 
surface pressure ($p_{\rm surf}$)
and the cross-track correction to the ephemeris ($\Delta \rho$).
The inner and outer green lines delimit the respective 1$\sigma$ and 3$\sigma$
error domains, see text for details.
{\it Right:} 
Evolution of Pluto's atmospheric pressure between 1988 and 2020.
All the error bars are displayed at 1$\sigma$ level.
Some point do not show error bars because they are smaller than the squares.
The black points are the values published by \cite{mez19} (M19) and
the red point is the result of the present work, as derived from the left panel.
The gray points are from other works. In chronological order, they are:
H17, the New Horizons value from the radio science data occultation 
(ingress point, 14 July 2015, \citealt{hin17}).
Note that this point is superimposed onto our own point of 29 June 2015;
Y21, the value derived by \cite{you21} from the ground-based, multi-chord 15 August 2018 occultation;
and
A20, the result of \cite{ari20} obtained from a grazing, single-chord occultation observed on 17 July 2019.
The colored lines are the modeled annual evolution of the surface pressure 
obtained with the Pluto volatile transport model described in \cite{mez19}.
The blue, green and red curves correspond to N$_2$ ice albedos of 0.73, 0.725 and 0.72,
respectively.
}%
\label{fig_t_p}
\end{figure*}

In Figure~\ref{fig_t_p}, we plot our measurement of $p_{\rm surf}$ in 2020 (red point)
along with other published values \citep{hin17,mez19,ari20,you21}.
%
The 6 June 2020 occultation shows that the pressure increase prevailing
between 1988 and 2013 stopped, and reached a stationary regime since 2015.
This is in line with the Pluto volatile transport model described in \cite{mez19}.

Our results do not support the rapid pressure decrease claimed by \cite{ari20} 
(who also used the ray-tracing method of \citealt{dia15})
from an occultation observed on 17 July 2019, see the point `A20' in Fig.~\ref{fig_t_p}.
With closest approach to Pluto's shadow center of 1008~km, the work of \cite{ari20} is based on 
an occultation that was more grazing than the event reported here (with closest approach of 735~km).
This induces a larger correlation between the parameters $p_{\rm surf}$ and $\Delta \rho$.
In particular, \cite{ari20} mention that the pressure drop they find between 2016 and 2019 
is actually detected at the 2.4$\sigma$ level, and thus remains marginally significant.
We thus estimate that the 2019 data point lacks accuracy to claim that a large
decrease occurred in 2019, followed by a return to a pressure close to that 
of 2015 during the year 2020 (this work).

We do not confirm either the decrease of pressure reported by \cite{you21}, 
based on the 15 August 2018 occultation observed from about two dozen stations 
in USA and Mexico, from which \cite{you21} derive a value 
$p_{\rm surf} = 11.13 \pm 0.4$~$\mu$bar (the point `Y21' in Fig.~\ref{fig_t_p}),
using a method that is not described by these authors.

It is important to note that all the points derived by us between 2002 and 2020 (\citealt{mez19} 
and present work) are obtained using a unique template temperature profile $T(r)$.
This assumption is backed up by the fact that although the pressure increased by a factor
of about three between 1988 and 2016 \citep{mez19}, the retrieved temperature profiles in 
1988, 2002, 2012 and 2015 \citep[respectively]{yel97,hin17,sic03,you08,dia15} 
are all similar, with a strong positive thermal gradient in the lower part of the atmosphere 
that peaks at $T \sim 110$~K near $r=1215$~km, 
followed by a roughly isothermal upper branch with a mild negative thermal gradient.
This globally fits the methane-thermostat model of \cite{yel89}, where
the upper-atmosphere temperature is robustly maintained near 100~K through the radiative properties of 
atmospheric CH4, almost independently of its abundance, while the lower part is forced by
heat conduction with the cold surface near 38~K.
So even if small systematic errors are introduced at this stage, a consistent comparison 
using a constant $T(r)$
between events is possible, so that general trends on pressure evolution can be monitored.   
Thus, at this stage, the methodologies used by both \cite{ari20} and \cite{you21} should be compared
with ours in some well-defined test cases to see if our approaches are consistent, 
and thus, fully comparable. 

Note that our results could be compared with those of New Horizons, 
derived from the radio occultation experiment (REX) in July 2015. 
Values of
$p_{\rm surf} = 12.8 \pm 0.7$ and $10.2 \pm 0.7$~$\mu$bar at entry and exit, respectively,
were obtained \citep{hin17}. 
The difference between the two values can be attributed to a 5~km difference 
in radius at the two locations probed by REX, the lower value at exit corresponding to
higher terrains on Pluto.
As the entry value probed a point over Sputnik Planitia, where sublimation of N$_2$ 
takes place, it should be more representative of $p_{\rm surf}$ than the exit value.
We see that our value of $p_{\rm surf}= 12.71 \pm 0.14$~$\mu$bar derived from 
the 29 June 2015 occultation \citep{mez19}
is in excellent agreement with the REX value of July 2015
(point `H17' in Fig.~\ref{fig_t_p}),
giving confidence that our approach not only provides general trends, but also
good estimates of $p_{\rm surf}$.

\section{Conclusions}

The 6 June 2020 stellar occultation allowed us to constrain Pluto's atmospheric evolution.
The surface pressure that we obtain, $p_{\rm surf}= 12.23^{+0.65}_{-0.38} $~$\mu$bar,
shows that Pluto's atmosphere has reached a plateau since mid-2015,
a result which is in line with the Pluto volatile transport model discussed in \cite{mez19}.
Our result does not support the drops of pressure reported by \cite{you21} and \cite{ari20}
in 2018 and 2019, respectively.
These inconsistencies call for careful comparisons between methodologies 
before any conclusions based on independent teams be drawn.

We note that if the model presented in \cite{mez19} is correct, and considering 
the typical error bars derived from occultations, it will 
be difficult to firmly confirm a pressure drop before 2025.
%
Meanwhile, observations should be organized whenever possible, 
as unaccounted processes may cause pressure changes not predicted by models.
%
\begin{acknowledgments}
The authors are grateful to the Directors of ARIES and IIA for granting time under Directors' Discretionary Time allotment for observing this event. The authors would also like to thank the staff of the Devasthal Observatory and the IR Group at TIFR for their help during observation. 
The work leading to these results has received funding from the 
European Research Council under the European Community's H2020
2014-2021 ERC Grant Agreement no. 669416 ``Lucky Star".
Research at PRL and IIST is supported by the Department of Space, Government of India.
M.A., G.B.-R., F.B.-R. and R.V.-M. thank CNPq and CAPES for grants 313144/2020-6,
314772/2020-0, 465376/2014-2 and Process 88887.310463/2018-00 - Project
88887.571156/2020-00.
%
%
R.V.-M. also thanks grant CNPq 304544/2017-6.
P.S-S. acknowledges financial support by the Spanish grant 
AYA-RTI2018-098657-J-I00 ``LEO-SBNAF" (MCIU/AEI/FEDER, UE).
\end{acknowledgments}
\vspace{5mm}
\facilities{%
Devasthal: 3.6-m, 1.3-m
}%

\end{document}